\documentstyle[12pt]{article}
\begin{document}
\begin{center}{\Large \bf Definition of fractal measures arising from 
fractional calculus}
\end{center}

\vspace{0.05in}
\centerline{Kiran M. Kolwankar\footnote{\it Department of Mathematics, Indian Institute
of Science, Bangalore 560 012, India} and Anil D. Gangal\footnote
{\it Department of Physics, University of Pune,
Pune 411 007, India}}

\vspace{0.15in}
The sets and curves of fractional dimension have been constructed
and found to be useful at number of places in science~\cite{Man}. 
They are used to model
various irregular phenomena.  
It is wellknown that the usual calculus is inadequate to
handle such structures and processes. 
Therefore a new calculus should be 
developed which incorporates fractals naturally.
Fractional calculus, which is a branch of mathematics
dealing with derivatives and integrals of fractional order,
is one such candidate. 
The relation between ordinary calculus and measures on $I\!\!R^n$ 
is wellknown.
For example, an $n$-fold integration gives an $n$-dimensional
volume. Also, the solution 
of $df/dx=1_{[0,x]}$,
where $1_{[0,x]}$ is an indicator function of $[0,x]$, gives 
length of the interval $[0,x]$~\cite{Kol}.
The aim of this paper is to arrive at a definition of a fractal measure
using the concepts from the fractional calculus. 
Here we shall restrict ourselves
to simple subsets of $[0,1]$ and more rigorous treatment will be given
elsewhere.

We first define a differential of fractional 
order $\alpha$ ($0\leq \alpha \leq 1$) as follows:
$d^{\alpha}x = {d^{-\alpha}1_{dx}(x) / {dx^{-\alpha}}}$ where
\begin{eqnarray}
{{d^qf(x)}\over{[d(x-a)]^q}}={1\over\Gamma(-q)}{\int_a^x{{f(y)}
\over{(x-y)^{q+1}}}}dy,\;\;\;\mbox{for}\;\;\; q < 0,\label{eq:def1}
\end{eqnarray}
is the Riemann-Liouville fractional integral~\cite{OS}.
Now we define a "fractal integral" by
$ _aI\!\!D^{-\alpha}_bf(x) = \int_a^b f(x) d^{\alpha}x$,
written in discrete form as,
$
_aI\!\!D_b^{-\alpha}f(x) =\lim_{N\rightarrow \infty}  
\sum_{i=0}^{N-1} f(x_i^*)
{d^{-\alpha}1_{dx_i} / {[d(x_{i+1}-x_i)]^{-\alpha}}},
$
where $[x_{i}, x_{i+1}]$, $i=0,...,N-1$, $x_0=a$ and $x_N=b$, 
provide a partition of
the interval [a,b] and $x_i^*$ is some suitably chosen point
of the subinterval $[x_{i},x_{i+1}]$. 
We now define the fractional
measure of a subset $A\cap[0,x]$ (assuming it to be measurable) as 
${\cal{F}}^{\alpha}(A\cap[0,x])= {_0I\!\!D_x^{-\alpha}1_A(x)}$.
Consider an example of a one-third Cantor set $C$ with
dimension $d = \ln(2)/\ln(3)$.
For this set $\cal{F}$ can be written as
${\cal{F}}^{\alpha}(C) = {_0I\!\!D^{-\alpha}_1 1_C(x)}$. 
Now we choose $x_i^*$ 
to be such that $1_C(x_i^*)$ is the maximum in that interval, then
\begin{eqnarray}
{\cal{F}}^{\alpha}(C)
&=& \lim_{N\rightarrow \infty} \sum_{i=0}^{N-1} F_C^i {(x_{i+1}-x_i)^{\alpha}\over 
\Gamma(\alpha+1)},   \label{eq:intsol}
\end{eqnarray}
where $F^i_C$ is a flag function which is 1 if a point of
set $C$ belongs to the interval $[x_i,x_{i+1}]$ and zero otherwise.
Clearly this measure is infinite if $\alpha < d$
and zero if $\alpha > d$. At $\alpha = d$
we have ${\cal{F}}^{d}(C) = 
{1/ \Gamma(d+1)}$ whereas the Hausdorff measure~\cite{Man}
 ${\cal{H}}^d(C)=[\Gamma(1/2)]^d /\Gamma(1+d/2)$.

Recently, a new quantity viz. local fractional derivative (LFD), was
defined~\cite{KG1} as
\begin{eqnarray}
I\!\!D^qf(y) =  {\lim_{x\rightarrow y}}
{{d^q[f(x)-f(y)]}
\over{[d(x-y)]^q}} \;\;\;0 < q \leq 1,\;\;\; x>y,\label{deflocg}
\end{eqnarray}
where the RHS uses
 Riemann-Liouville fractional derivative~\cite{OS} given by
\begin{eqnarray}
{{d^qf(x)}\over{[d(x-a)]^q}}={1\over\Gamma(1-q)}{d\over{dx}}
{\int_a^x{{f(y)}
\over{(x-y)^{q}}}}dy\;\;\;\mbox{for}\;\;\; 0<q<1.\label{eq:def2}
\end{eqnarray}
We also introduced~\cite{KG4} 
local fractional differential equations which
involve LFDs. 
A solution of the equation~\cite{KG4}
$I\!\!D^\alpha f(x) = 1_{C}(x)$
turns out to be equation (\ref{eq:intsol})
implying $f(x)={\cal{F}}^{\alpha}(C\cap[0,x])$~\cite{Kol}. 
This generalizes
the fact that the solution of $f'(x) = 1_{[0,x]}$ is the
length of the interval $[0,x]$.

A local fractional diffusion equation
given by 
$I\!\!D_t^\alpha W(x,t)\! =\! {(1_C(t)/ 2)}$  
${(\partial^2W(x,t)/{\partial x^2})}
$ (compare Ref.~\cite{KG4}),
where $W(x,t)$ is a probability density for finding a particle 
in neighbourhood of $x$ at time $t$, has a solution
given by~\cite{KG4}, for $W(x,0) = \delta(x)$,
\begin{eqnarray}
W(x,t) &=& {1\over\sqrt{2\pi {\cal{F}}(C\cap [0,t]) }}
\exp({ - x^2\over{2{\cal{F}}(C\cap [0,t])}}).  \label{eq:soldiffusn}
\end{eqnarray}
The mean square displacement, 
$<x^2> = 2{\cal{F}}(C\cap [0,t])$, is proportional to 
$t^{\alpha}$. Hence
the equation~(\ref{eq:soldiffusn}) gives a subdiffusive solution.

We have introduced a definition of fractal measures using
fractional calculus and shown it to be useful in studying diffusion
in fractal time.

One of the authors (KMK) would like to thank DST (India)
(DST: PAM: GR: 381) for financial 
assistance.


\begin{thebibliography}{abc}
\bibitem{Man} Mandelbrot B. B., {\it The Fractal Geometry of Nature}
(Freeman, New York, 1977).
\bibitem{Kol} K. M. Kolwankar, Ph. D. thesis, University of Pune
(1997). (chao-dyn/9811008)
\bibitem{OS} K. B. Oldham  and J. Spanier,   
{\it The Fractional Calculus}
 (Academic Press, New York, 1974).
\bibitem{KG1} K. M. Kolwankar and A. D. Gangal, {\it Chaos} {\bf 6},
505 (1996). (chao-dyn/9609016)
\bibitem{KG4} K. M. Kolwankar and A. D. Gangal, 
{\it Phys. Rev. Lett.} {\bf 80} 214 (1998). (cond-mat/9801138)
\end{thebibliography}
\end{document}